\documentclass[aps,pre,twocolumn,superscriptaddress]{revtex4}
\usepackage{graphicx}
\usepackage{color}
\usepackage{amsmath}

\begin{document}

\title{Brownian motion under non-instantaneous resetting in higher dimensions}
\author{Anna S. Bodrova}
\affiliation{Humboldt University, Department of Physics, Newtonstrasse 15, 12489 Berlin, Germany}
\author{Igor M. Sokolov}
\affiliation{Humboldt University, Department of Physics, Newtonstrasse 15, 12489 Berlin, Germany}
\affiliation{IRIS Adlershof, Zum Gro{\ss}en Windkanal 6, 12489 Berlin, Germany}

\begin{abstract}
We consider Brownian motion under resetting in higher dimensions for the case when the return of the 
particle to the origin occurs at a constant speed. We investigate the behavior of the probability density 
function (PDF) and of the mean-squared displacement (MSD) in this process. We study two different resetting protocols: exponentially distributed time intervals between the resetting events (Poissonian resetting) and resetting at fixed time intervals (deterministic
resetting). We moreover discuss a general problem of the invariance of the PDF with respect to 
the return speed, as observed in the one-dimensional system for Poissonian resetting, and show, that
this one dimensional situation is the only one in which such an invariance can be found. However, the invariance of the MSD can still be observed in higher dimensions.
\end{abstract}

\maketitle

\section{Introduction} 

The investigation of different aspects of the behavior of random processes under resetting has attracted 
much attention in the last years, see \cite{review} for a recent review. In such processes the stochastic motion is interrupted from time to time and restarts anew. These processes may be applicable in order to describe computer search algorithms, motion of animals returning home to rest and of machines returning in order to recharge.

The very first model studied, see \cite{EvansMajumdar}, corresponded to a one-dimensional Brownian motion interrupted by the resetting events following a Poisson process. While the probability density function (PDF) of the unrestricted Brownian motion spreads with time, the PDF of diffusion with resetting reaches,  after a short relaxation period, a stationary state, which corresponds to a Laplace distribution of the coordinate \cite{EvansMajumdar}. Later, Brownian motion in two and higher dimensions has been studied \cite{high12,bhat}. Other types of processes under resetting, such as L\'evy flights (LF) \cite{levy1,levy2}, fractional diffusion \cite{santos}, continuous-time random walks (CTRW) with or without drift \cite{MV2013,MC2016,Sh2017,ctrw,ctrwour}, random acceleration process \cite{rap}, and scaled Brownian motion (SBM) \cite{Anna01, Anna02} have been also considered. 
Also a variety of waiting time distributions, such as $\delta$-distribution (resetting at a fixed time after starting the stochastic motion) \cite{shlomi2017, palrt}, power-law \cite{NagarGupta,Anna01, Anna02}, and other types \cite{res2016,shlomi2016} have been investigated.

The return to the initial position (assumed at the origin of the coordinate system in what follows) is treated as instantaneous in the classical approach \cite{review}. This means that at the resetting event the particle teleports to the initial position. While this model is immediately applicable to computer algorithms \cite{computerscience}, it is unrealistic with respect to the motion of foraging animals, robots and other objects. Recently, the diffusion under resetting with non-instantaneous return has been considered, but for one-dimensional systems only \cite{shlomi,shlomi1,shlomi2,me,campos}. Different types of the return motion can be studied: the return at constant velocity, at constant acceleration and with constant return time \cite{me}. The first case is the most studied one and corresponds to the resetting by constant force in the over-damped limit. It can be realized experimentally for the motion of colloidal particles in the liquids under the action of holographic optical tweezers \cite{exp}. This type of return will be also considered in the current study. The second case is the underdamped return motion under the action of the constant force \cite{me}. The third type is the return under the action of the harmonic force and can be also easily implemented in the experiments with resetting of colloidal particles via the holographic optical tweezers \cite{exp}.

Surprisingly, it has been revealed, that the PDF and MSD of normal diffusion under Poisson resetting do not depend on the return velocity, if it remains constant during the whole return path \cite{shlomi1,shlomi2,campos,me}. However, this invariance does not hold for non-Poissonian resetting, such as resetting at fixed time \cite{me} and for resetting with Pareto distribution of the waiting times \cite{campos}.


The motion of real objects, e.g. of a robotic vacuum cleaner performing stochastic motion during the cleaning phase and returning to its base in order to recharge, as well as of foraging mammals, is usually two-dimensional (2d); the motion of birds, fish and drones is quasi three-dimensional (3d). Recently the two-dimensional diffusion of colloidal particle under resetting via holographic optical tweezers has been realized in the experiments \cite{exp, exp2}. This gave us motivation to study Brownian motion under resetting with non-instantaneous return in higher dimensions. 

In the current study we generalize the approach, developed in \cite{me}. We consider MSD and PDF of the particle performing Brownian motion with non-instanteneous return at constant velocity. We investigate if the invariance of the PDF and MSD with respect to the return velocity can be observed also in higher dimensions and proceed as follows. In Section II we describe our model and give the general expressions for MSD and for PDF. In Section III we explicitly study the PDF and MSD of normal diffusion under Poisson and deterministic non-instantaneous resetting in 2d and 3d in terms of numerical simulations and analytically. In Section IV we address the invariance of the PDF and MSD with respect to the return velocity. Finally, concluding remarks are given in Section V.

\section{Model}

We study Brownian motion under non-instantaneous resetting in higher dimensions by generalizing the approach used in Ref~\cite{me} for the one-dimensional case. We assume that the object (particle) starts its motion at the origin $\mathbf{R}=0$. The motion of the particle is a renewal process consisting of repeated runs.
Each run is a sequence of two processes (phases of motion): The
stochastic displacement process which ends at the resetting event, and the deterministic return
process, which ends when the particle returns to the origin $\mathbf{R}=0$. 
We assume that the motion is statistically isotropic (i.e. mirror, rotationally or spherically symmetric in dimensions 1, 2, and 3 correspondingly), so that the PDF of displacements is determined only by the distance $R$ from the origin. The resetting occurs with waiting time density $\psi(t)$, which is taken to be either exponential or deterministic with fixed time interval. 

\subsection{Poisson resetting}

Poisson resetting occurs with a constant rate. It means that the probability for the event to happen is the same at any time moment. The waiting time distribution for the resetting event in this case is exponential:
\begin{equation}\label{pexp}
\psi(t)=r\exp(-rt) 
\end{equation}
with $r$ being the rate of the resetting events. 
The survival probability \cite{sokbook}
\begin{equation}\label{psurv}
\Psi(t)=1-\int_0^t\psi(t^{\prime})dt^{\prime}
\end{equation}
is the probability that no resetting event occurred up to time $t$. For Poisson resetting it is equal to 
\begin{equation}\label{survexp}
\Psi(t)=e^{- rt}.
\end{equation}
The mean resetting time is 
\begin{equation} \label{tres}
\langle t_{\rm res}\rangle=\int_0^{\infty}dt\psi(t)t=\frac{1}{r}
\end{equation}
Other moments are 
\begin{eqnarray}
\langle t_{\mathrm{res}}^2 \rangle &=& \frac{2}{r^2}\\
\langle t_{\mathrm{res}}^{\frac12} \rangle &=&  \frac{\sqrt{\pi}}{2\sqrt{r}}\\
\langle t_{\mathrm{res}}^{\frac32} \rangle &=&  \frac{3\sqrt{\pi}}{4r\sqrt{r}}.\label{tres3}
\end{eqnarray}

\subsection{Deterministic resetting}

 The search process is most effective in the case of the deterministic resetting, occurring with constant time intervals between the resetting events with waiting time distribution \cite{shlomi2017,palrt,sokprl}
\begin{equation}\label{psidelta}
\psi(t)=\delta(t-t_r).
\end{equation}
The survival probability (Eq.~\ref{psurv}) for resetting events 
is now given by the Heaviside function $\Psi(t)=\Theta(t_r-t)$. The mean duration of the displacement phase is 
\begin{equation}\label{tresdet}
\langle t_{\mathrm{res}} \rangle =t_r
\end{equation}
and the other moments are
\begin{equation}\label{tresdeta}
\langle t_{\mathrm{res}}^{\alpha} \rangle =t_r^{\alpha}\,.
\end{equation}

\subsection{Probability density function}
The stationary PDF follows by noting that a time at which the particle's position is observed may fall into the 
displacement phase of the motion or into the return phase, and these events are mutually exclusive \cite{me}. The probability 
of the first, $P_1$, and of the second, $P_2$, are given by 
\begin{equation}
 P_1= \frac{\langle t_{\rm res}\rangle}{\langle t_{\rm res}\rangle+\langle t_{\rm ret} \rangle}, \qquad  P_2= \frac{\langle t_{\rm ret}\rangle}{\langle t_{\rm res}\rangle+\langle t_{\rm ret} \rangle},
\end{equation}
where $\langle t_{\rm res}\rangle$ is the mean duration of the displacement phase, $\langle t_{\rm ret}\rangle$
is the mean duration of the return phase, and 
\begin{equation}
 \langle t_{\rm res}\rangle+\langle t_{\rm ret} \rangle = \langle t_{\rm run}\rangle
\end{equation}
represents the mean duration of a run.  The PDF of the particle's positions can thus be represented as
\begin{equation}
p(R)=p_1(R) P_1 + p_2(R)P_2,
\label{eq:Cond}
\end{equation}
where $p_1(R)$ and $p_2(R)$  are the PDFs of the particle's displacements from the origin conditioned on the fact that the time at which the position was measured belongs to the displacement phase and to the return phase of the motion  correspondingly. Let us now introduce the rescaled PDF in the displacement phase $\rho_1(R)=p_1(R)P_1\langle t_{\rm run}\rangle$ and the rescaled PDF of the return phase $\rho_2(R)=p_2(R)P_2\langle t_{\rm run}\rangle$. According to Ref. \cite{me} these are given by
\begin{equation}\label{I1}
 \rho_1(R)=\int_0^\infty dt_{\mathrm{res}} \psi(t_{\mathrm{res}}) \int_0^{t_{\mathrm{res}}} P(R|t) dt 
\end{equation}
under general conditions, and 
\begin{equation}  \label{I2v}
\rho_2(R)=\frac{1}{v}\int_{R}^\infty dR^{\,\prime} \int_0^\infty dt\; P(R^{\,\prime}|t) \psi(t) 
\end{equation}
for the return at a constant speed $v$. Here $P(R|t)$ is the propagator of free motion. In the current manuscript we investigate free Brownian motion (see the next subsection D), but also other types of motion, such as CTRW, LF and SBM can be considered. The mean return time is then
\begin{equation}\label{tretI}
\langle t_{\mathrm{ret}} \rangle=\int_0^{\infty}\rho_2(R)dR=\frac{1}{v} \int_0^\infty dt \psi(t) \int_0^\infty dR' R' P(R'|t)\,.
\end{equation}
Consequently, Eq.(\ref{eq:Cond}) then takes the form 
\begin{equation}\label{mixture}
 p(R) = \frac{\rho_1(R) + \rho_2(R)}{\langle t_{\rm res}\rangle+\langle t_{\rm ret} \rangle}.
\end{equation}

\begin{figure*}[ht]
\centerline{\includegraphics[width=0.98\columnwidth]{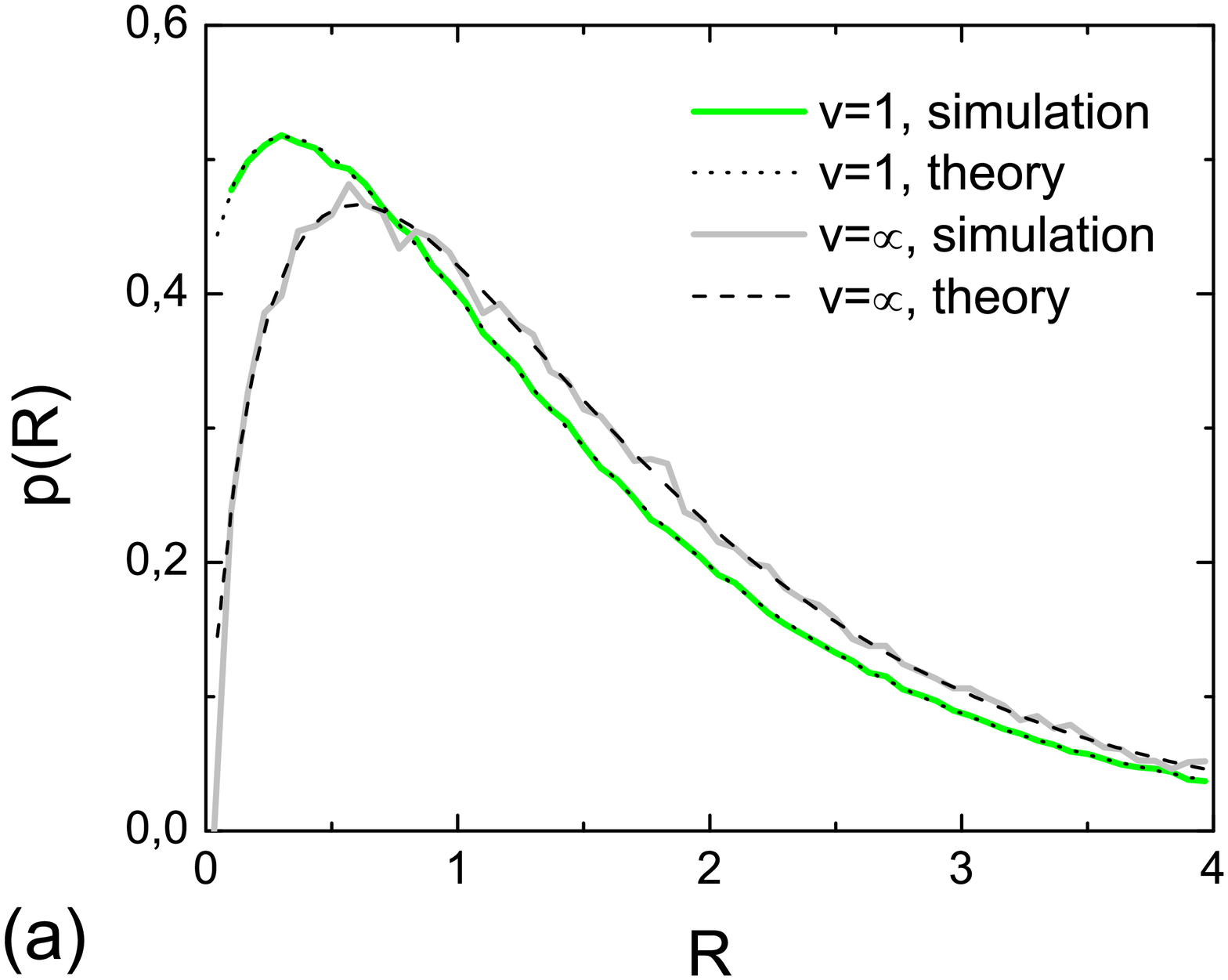}\includegraphics[width=0.98\columnwidth]{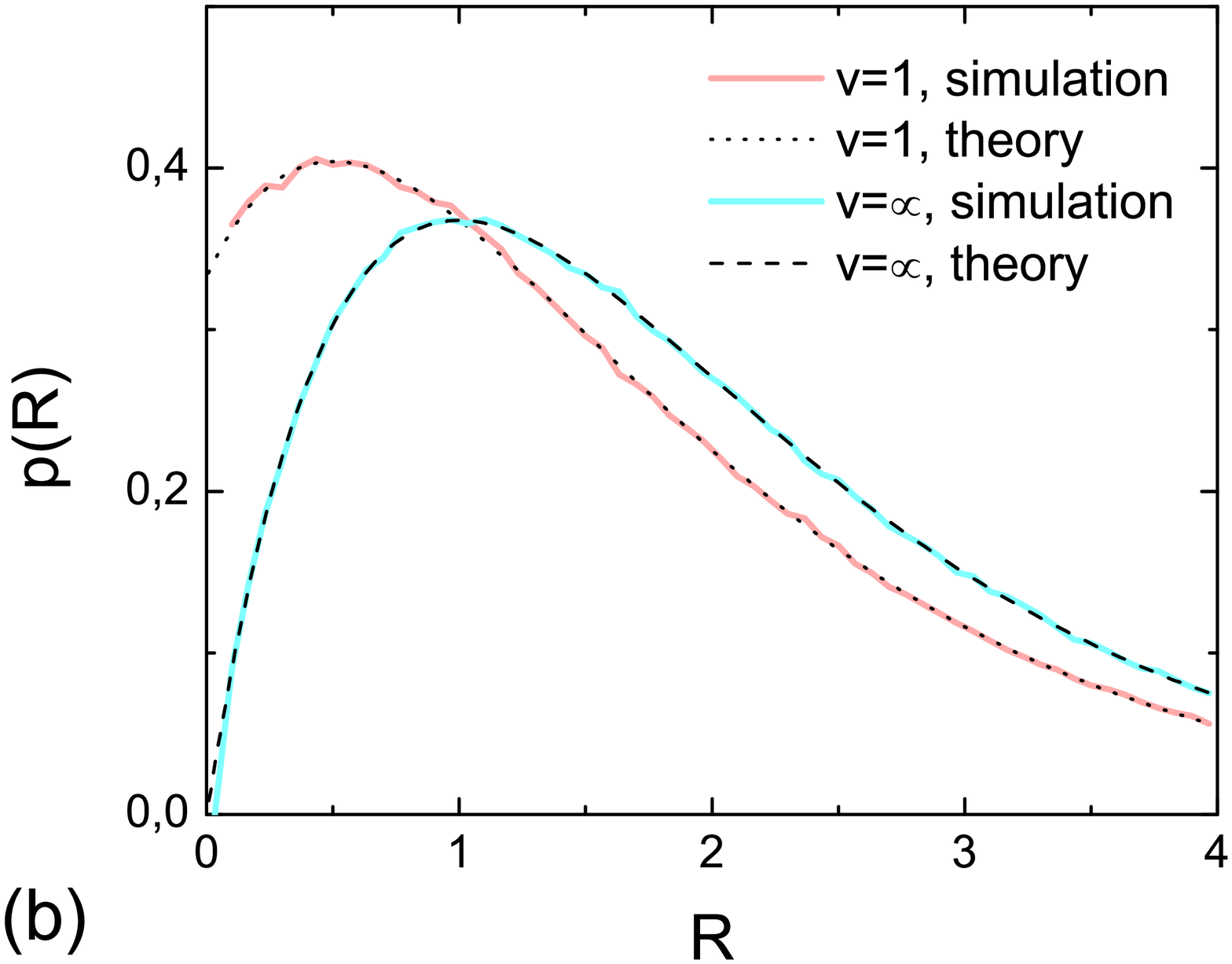}}
\caption{Probability density function of the distance of the particle to the origin $p(R)$ for the exponential resetting in (a) 2d and (b) 3d with $D=1$, $r=1$. The colored solid lines show the results of numerical simulations, the thin black dashed and dotted lines show PDF given by Eqs.~(\ref{pdfexp2d}) and (\ref{pdfexp3d}) for 2d and 3d, correspondingly. For comparison we provide also the results for the instantaneous resetting ($v=\infty$).}
\label{Gexp}
\end{figure*} 

\subsection{Brownian motion in displacement phase}
The propagator of free Brownian particle 
in $d$ dimensions reads
\begin{equation}\label{pgau}
P(R|t)=\frac{dR^{d-1}}{\left(4D t\right)^{d/2}\Gamma\left(1+\frac{d}{2}\right)} e^{- \frac{R^2}{4 D t}}\,,\end{equation}
where $D$ is the diffusion coefficient. Using Eq.~(\ref{tretI}) we get for the ballistic return time 
\begin{equation}
 \langle t_{\mathrm{ret}}\rangle =\frac{1}{v}\lambda(d)D^{1/2}\langle t^{1/2}_{\mathrm{res}}\rangle
\end{equation}
with the coefficients
\begin{equation}\label{lambda}
\lambda(d) =2\frac{\Gamma\left(\frac{d+1}{2}\right)}{\Gamma\left(\frac{d}{2}\right)}\,.
\end{equation}

 The general expression for probability density function in higher dimensions can be calculated using Eqs.~(\ref{I1}) - (\ref{mixture}) with the propagator given by Eq.~(\ref{pgau}):
\begin{equation}\label{rhomainsuper}
\rho(R) =  \frac{\int_0^{\infty}dt^{\prime}\psi(t^{\prime})\left(\frac{R}{2D}\Gamma\left(\frac{d}{2}-1;\frac{R^2}{4Dt^{\prime}}\right)+\frac{1}{v}\Gamma\left(\frac{d}{2};\frac{R^2}{4Dt^{\prime}}\right)\right)}{ \Gamma\left(\frac{d}{2}\right)\left(\langle t_{\rm res}\rangle + \frac{1}{v}\sqrt{D}\lambda(d)\langle t_{\mathrm{res}}^{\frac12}\right) \rangle }\,.
\end{equation}

For the Poisson resetting with waiting time distribution, given by Eq.~(\ref{pexp}), the integration yields
\begin{equation}\label{rhomainexp}
\rho(R) = \frac{2}{\Gamma\left(\frac{d}{2}\right)}\!\left(\frac{rR^2}{4D}\right)^{\frac{d}{4}}\!\!\frac{\frac{1}{\sqrt{Dr}}K_{1-\frac{d}{2}}\left(R\sqrt{\frac{r}{D}}\right)+\frac{1}{v}K_{\frac{d}{2}}\left(R\sqrt{\frac{r}{D}}\right)}{\frac{1}{r}+\frac{\sqrt{\pi D}}{2v\sqrt{r}}\lambda(d)}\,.
\end{equation}
and for the deterministic resetting (Eq.~\ref{psidelta})
\begin{equation}\label{rhomaindet}
\rho(R) = \frac{\frac{R}{2D}\Gamma\left(\frac{d}{2}-1;\frac{R^2}{4Dt_r}\right)+\frac{1}{v}\Gamma\left(\frac{d}{2};\frac{R^2}{4Dt_r}\right)}{ t_r\Gamma\left(\frac{d}{2}\right) + \frac{1}{v}2\sqrt{Dt_r}\Gamma\left(\frac{d+1}{2}\right) }\,.
\end{equation}

\subsection{Mean squared displacement}
The MSD of free Brownian motion under non-instantaneous return at constant velocity in the steady state
\begin{equation}\label{msdpdf}
\left\langle R^2\right\rangle = \int_0^{\infty} R'^2 p(R^{\prime})dR^{\prime}\,.
\end{equation}
may be obtained by inserting Eq.~(\ref{rhomainsuper}) into Eq.~(\ref{msdpdf}) and performing the integration
\begin{equation}\label{msdmain}
\langle R^2 \rangle = D\frac{ d \langle t^2_{\mathrm{res}}\rangle + \frac{1}{v}\kappa(d)D^{1/2} \langle t^{3/2}_{\mathrm{res}}\rangle}{\langle t_{\mathrm{res}}\rangle + \frac{1}{v} \lambda(d)D^{1/2}\langle t^{1/2}_{\mathrm{res}}\rangle}\,.
\end{equation}
Here the coefficients $\lambda(d)$ are given in terms of Eq.~(\ref{lambda}) and the coefficients $\kappa(d)$ have the form
\begin{equation}\label{kappa}
\kappa(d)=\frac{8\Gamma\left(\frac{3+d}{2}\right)}{3\Gamma\left(\frac{d}{2}\right)}\,.
\end{equation}
For the Poisson resetting we introduce the time moments, Eqs.~(\ref{tres}-\ref{tres3}), into Eq.~(\ref{msdmain})
\begin{equation}\label{msdmainexp}
\langle R^2 \rangle = \frac{d D}{r}\frac{ 2 + \frac{3\kappa(d)\sqrt{\pi rD}}{4dv} }{1 + \frac{\lambda(d)\sqrt{\pi rD}}{2v} }\,.
\end{equation}
In the case of the deterministic resetting we use Eqs.~(\ref{tresdet}-\ref{tresdeta}) and get
\begin{equation}\label{msdmaindet}
\langle R^2 \rangle = Dt_r\frac{ dt_r + \frac{1}{v}\kappa(d)\sqrt{D t_r}}{ t_r + \frac{1}{v} \lambda(d) \sqrt{Dt_r}}\,.
\end{equation}

\begin{figure*}[ht]
\centerline{\includegraphics[width=0.98\columnwidth]{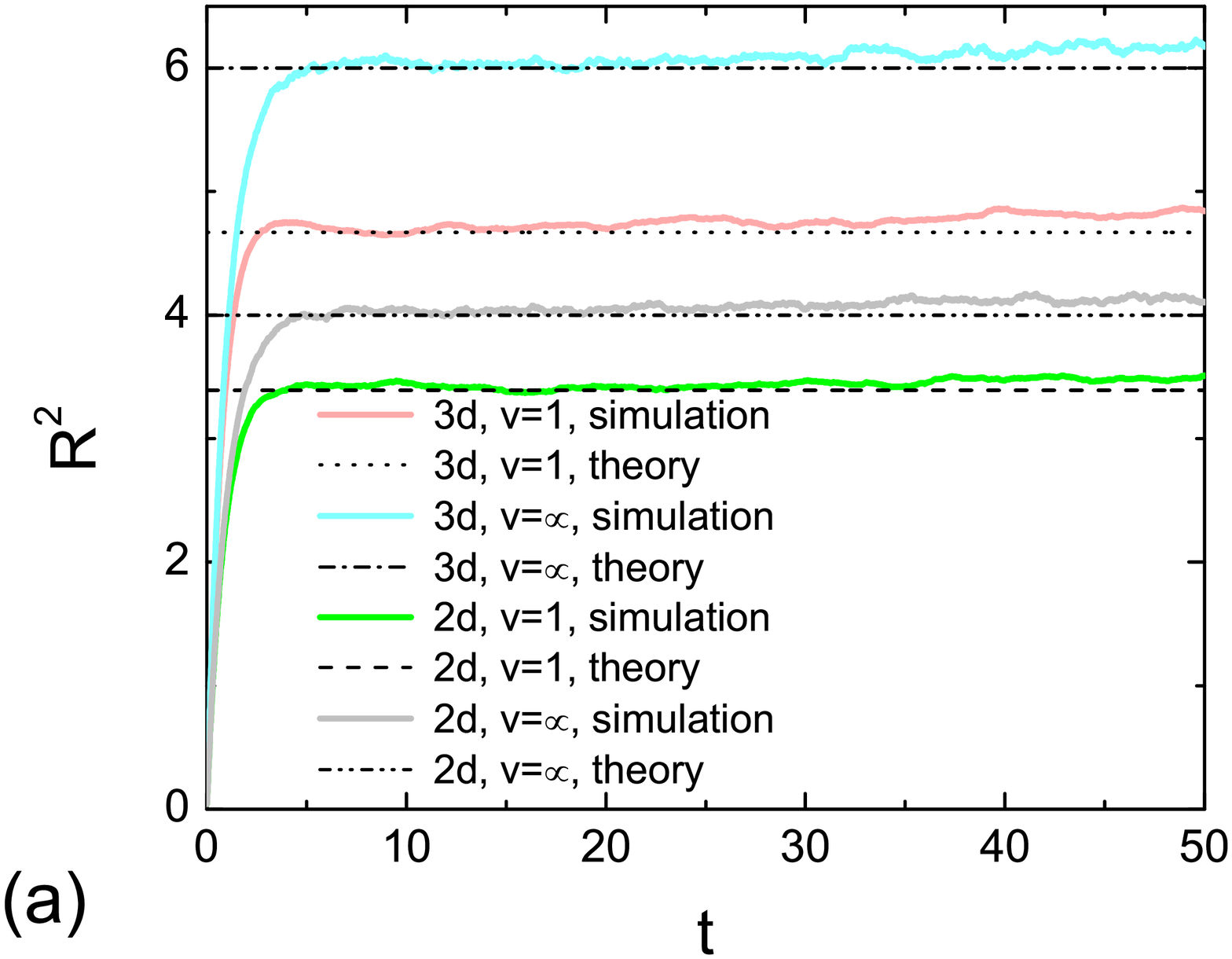}\includegraphics[width=0.98\columnwidth]{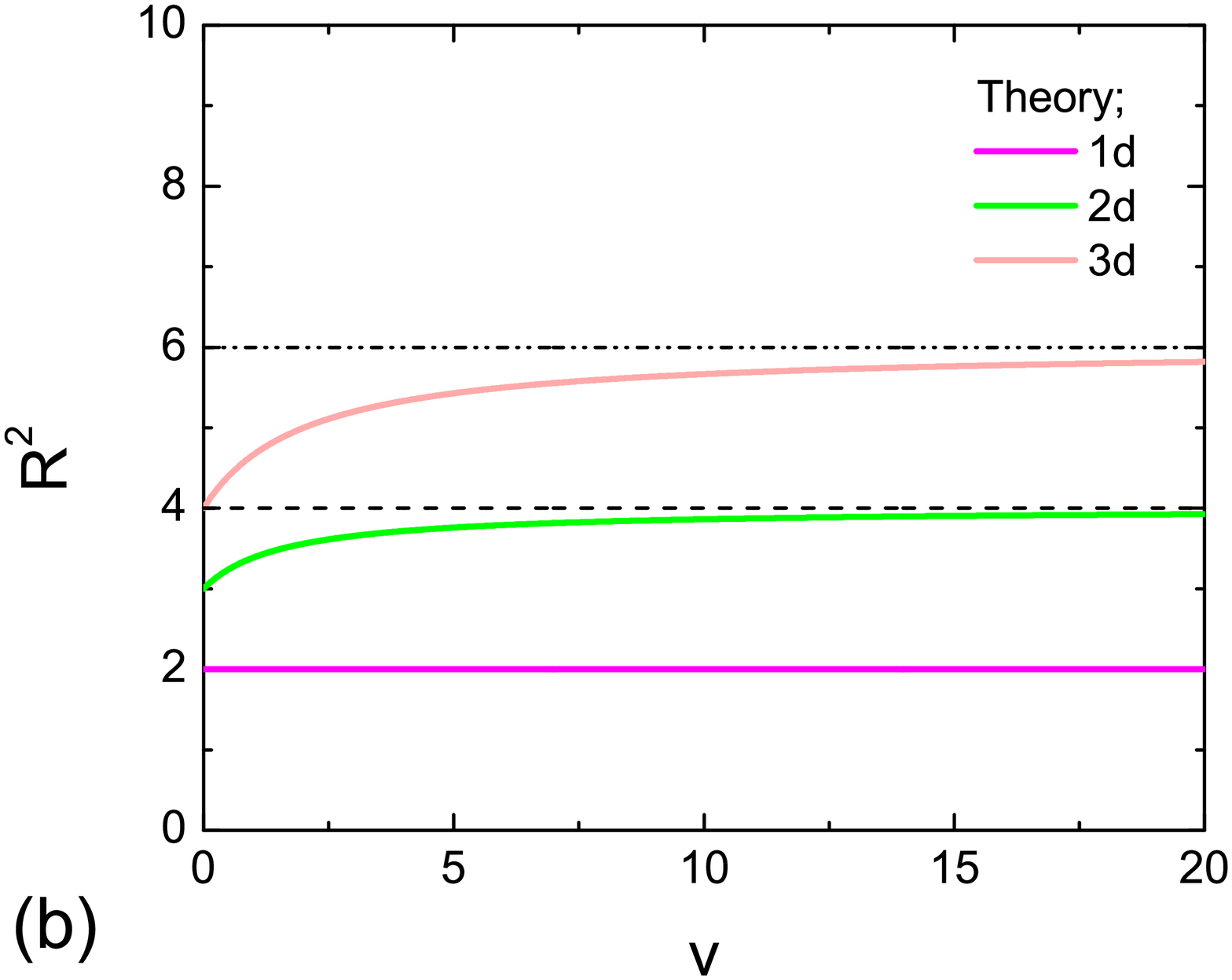}}
\caption{(a) Mean squared displacement for the exponential resetting ($D=1$, $r=1$) as functions of time (simulations), 
showing the relaxation to stationary values (horizonlal lines) given by Eqs.~(\ref{R2expd2}) and (\ref{R2expd3}). (b) Stationary MSDs as functions of the return velocity given by Eqs.~(\ref{x2exp}), (\ref{R2expd2}) and (\ref{R2expd3}) in $1d$, $2d$ and $3d$, correspondingly. Black horizontal lines represent the limiting values for $v \to \infty$ in $d=2,3$.}
\label{GR2exp}
\end{figure*} 

\section{Results and discussion}

\subsection{Computer simulations}
We compare the analytical predictions for PDF (\ref{rhomainexp}-\ref{rhomaindet}) and MSD (\ref{msdmainexp}-\ref{msdmaindet}) with numerical simulations in one, two and three-dimensional systems. We use the discretization of Langevin equations in the motion phase and of the equation 
of motion for the return at a constant speed. The time axis is discretized with the step $dt=t_{i+1}-t_i$, and the time of 
the first resetting event is generated according to its probability density $\psi(t)$. For the deterministic resetting the 
resetting time is fixed. During the displacement phase the particle performs stochastic motion according to a finite-difference 
analogue of the Langevin equation in 3d
\begin{eqnarray}\label{lattice}
x_{i+1}=x_i+\xi_{ix}\sqrt{2Ddt}\\
y_{i+1}=y_i+\xi_{iy}\sqrt{2Ddt}\\
z_{i+1}=z_i+\xi_{iz}\sqrt{2Ddt}
\end{eqnarray}
Here $x_i=x(t_i)$, $y_i$ and $z_i$ are the coordinates of the particle at time $t_i$, $R^2=x_i^2+y_i^2+z_i^2$ and
$\xi_{ix,y,z}$ are the random numbers undergoing a standard normal distribution and generated using the Box-Muller transform. In 2d $z_i=0$ and is not updated during the simulation. In 1d both $z_i=0$ and $y_i=0$.
When the resetting event occurs, the particle starts moving to the origin at a constant speed: $R_{i+1}=R_i - v dt$. 
When the particle reaches the origin, the time of the next resetting event is generated, and the particle starts performing Brownian motion again until the new resetting event. 
All simulations are performed with $N=10^5-10^6$ particles.

\subsection{Exponential resetting}

\subsubsection{One dimension}

Here we briefly review the results obtained for the Brownian motion with Poisson resetting in 1d. In this case the remarkable fact of the invariance of the PDF and MSD with respect to the return speed has been observed \cite{shlomi1,shlomi2,me,campos}. The PDF is given by the Laplace distribution and has the same form as for the Brownian motion under instantaneous resetting \cite{EvansMajumdar};
\begin{equation}
 p(R)=  \sqrt{\frac{r}{D}} \exp\left(-\sqrt{\frac{r}{D}}R\right)\,. \label{pdfexp}
 \end{equation}
 The MSD is also invariant with respect to the return speed and is given by
 \begin{equation}\label{x2exp}
\left\langle R^2\right\rangle = \frac{2D}{r}\,.
\end{equation}

 \begin{figure*}[ht]
\centerline{\includegraphics[width=0.98\columnwidth]{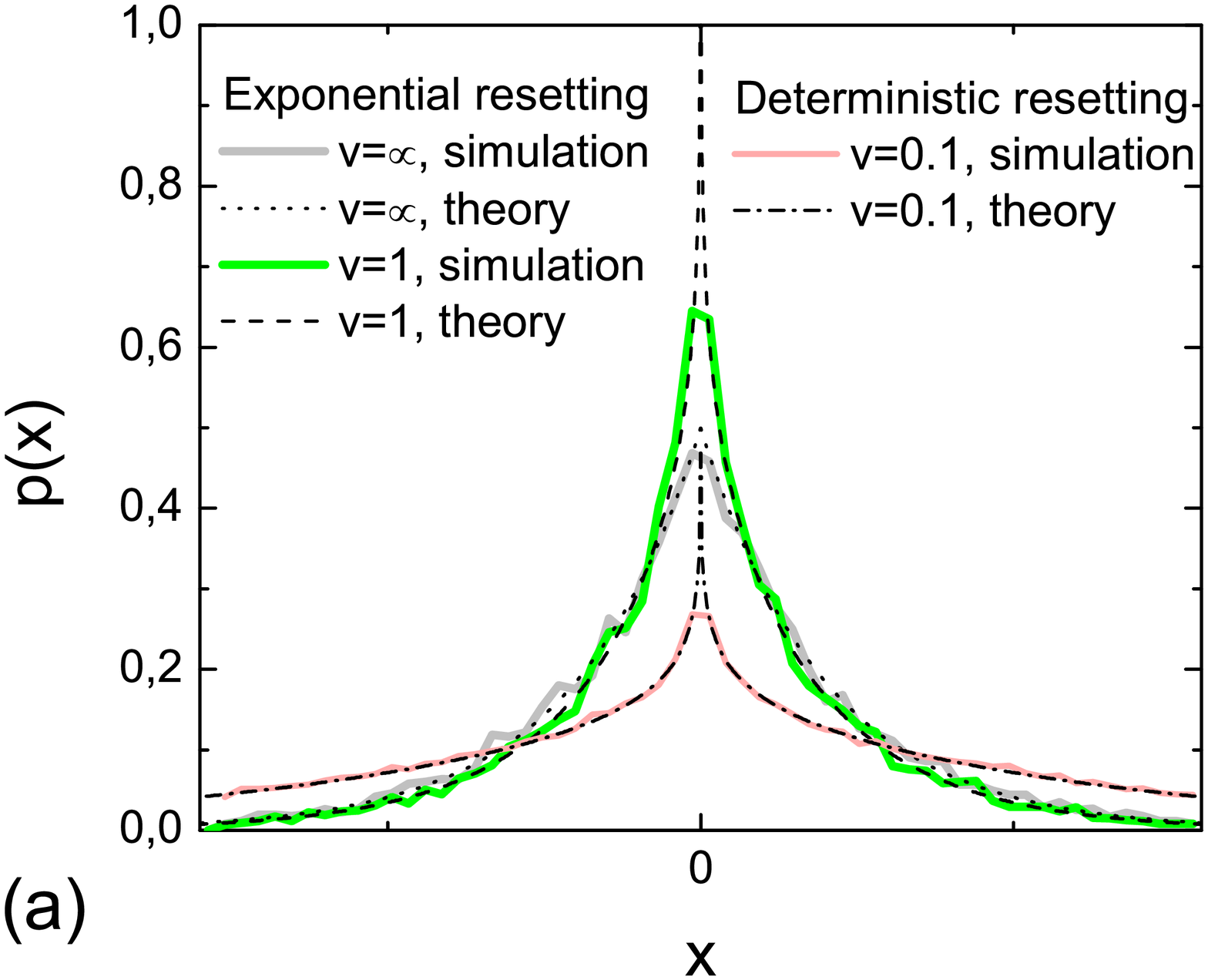}\includegraphics[width=0.98\columnwidth]{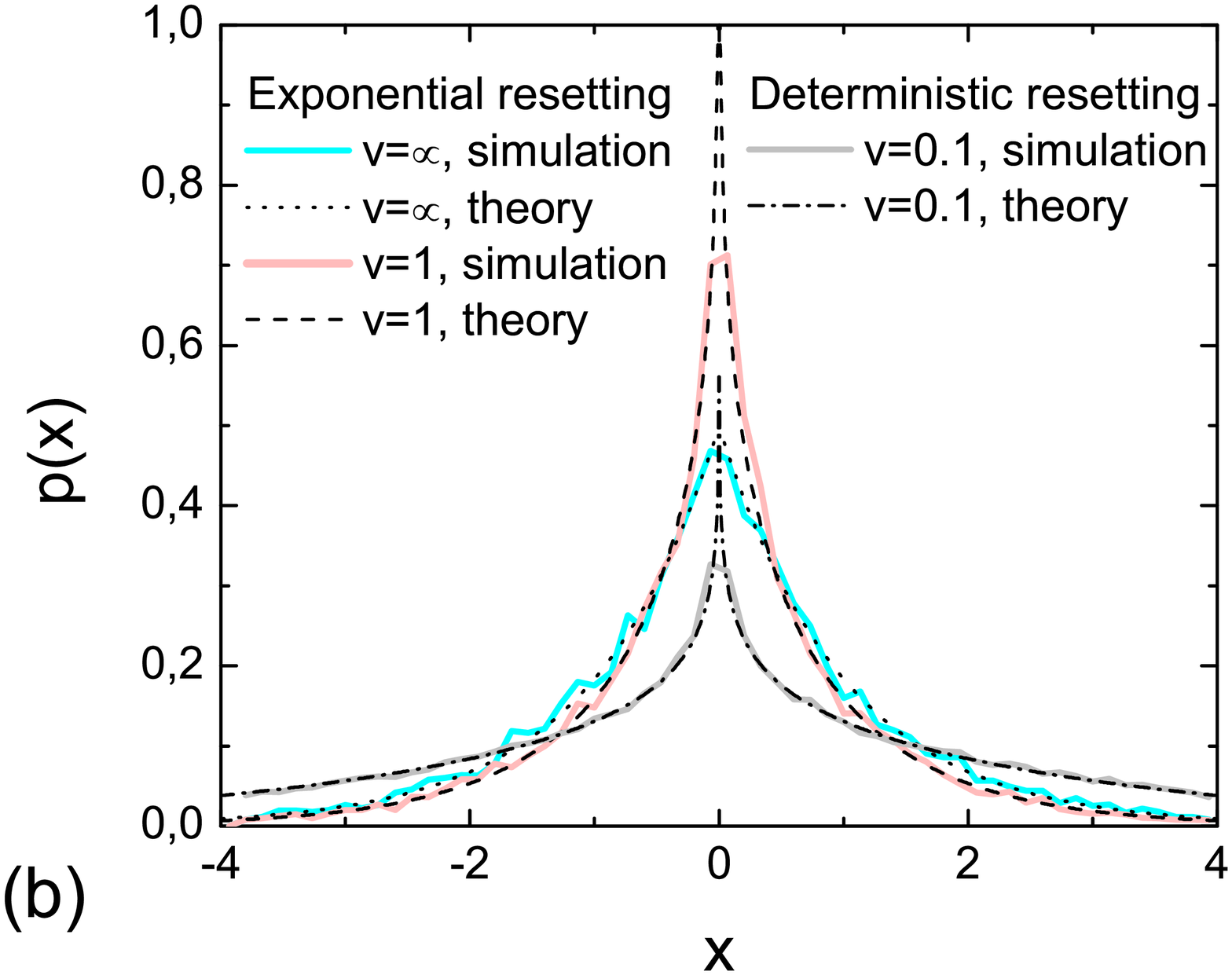}}
\caption{Probability density function $p(x)$ of the projection $x$ for exponential and deterministic resetting in (a) 2d and (b) 3d. The colored solid lines show the results of numerical simulations, the thin black dashed and dotted lines show analytical result (Eqs.~(\ref{pdfx2d}) and (\ref{pdfx3d})) for exponential resetting with $D=1$, $r=1$ and (Eqs.~(\ref{pdfxdet2d}) and (\ref{pdfx3det})) for deterministic resetting with $t_r=10$ in 2d and 3d, correspondingly.}
\label{Gpdf1d}
\end{figure*} 

\subsubsection{Two dimensions}
The PDF calculated according to Eq.~(\ref{rhomainexp}) with $d=2$ is
\begin{equation}\label{pdfexp2d}
p(R)=\frac{\frac{R}{D}K_0\left(R\sqrt{\frac{r}{D}}\right)+\frac{1}{v}R\sqrt{\frac{r}{D}}K_1\left(R\sqrt{\frac{r}{D}}\right)}{\frac{1}{r}+\frac{\pi}{2v}\sqrt{\frac{D}{r}}}\,.
\end{equation}
Here $K_\alpha(R)$ is the modified Bessel function of second kind. The same result has been obtained independently by Tal-Friedman et al. \cite{exp}. The PDF given by Eq.~(\ref{pdfexp2d}) is compared with numerical simulations in Fig.~\ref{Gexp}, and a very good agreement is found. The whole distribution becomes broader and its mode shifts closer to the origin for smaller return velocities.

The MSD can be obtained by introducing $d=2$ into Eq.~(\ref{msdmainexp}):
\begin{equation}\label{R2expd2}
\left\langle R^2\right\rangle = \frac{D}{r}\frac{8v+3\pi\sqrt{Dr}}{2v+\pi\sqrt{Dr}}\,.
\end{equation}

 According to Eqs.~(\ref{pdfexp2d}-\ref{R2expd2}) the invariance of the PDF and MSD with respect to the return 
 speed found in one dimensional systems does not hold anymore in two dimensions: the PDF and the MSD depend explicitly on the speed of the return motion. Numerical simulations reveal that the MSD rapidly reaches its steady state value given by Eq.~(\ref{R2expd2}) in 2d (Fig.~\ref{GR2exp}a). This steady state grows with increasing of the return velocity $v$ and tends to $\langle R^2 \rangle=4 D/r$ (Fig.~\ref{GR2exp}b).
 
Now let us consider the probability density of the projection of the particle's position of the particle on an arbitrary axis. Due to isotropy of the system it is equal to probability density $p_x(x)$ of the $x$-coordinate. 
For instantaneous resetting the coordinates $x=R\cos\phi$ and $y=R\sin\phi$ evolve independently of each other, and $p_x(x)$ is the same as the 1d distribution obtained by Evans and Majumdar (Eq.~\ref{pdfexp}) \cite{EvansMajumdar}. In the case of the finite return velocity the evolution of the coordinates is coupled, because the duration of the return along the $x$-coordinate depends on the $y$-coordinate as well: the larger $y$, the longer the return. The PDF $p_x(x)$ may be derived from $p(R)$ given by Eq.~(\ref{pdfexp2d}) in the following way:
\begin{equation}
 p_x(x)= \int_0^{\infty}dR\int_0^{2\pi}d\phi \delta\left(x-R\cos\phi\right)p(R)p(\phi).
 \end{equation}
Due to the isotropy of our system $p(\phi)=\frac{1}{2\pi}$, and the integration over $\phi$ yields
\begin{equation}
 p_x(x)= \frac{1}{\pi}\int_{|x|}^{\infty}dR \frac{p(R)}{\sqrt{R^2-x^2}}. \label{pdfx}
 \end{equation}
Substituting Eq.~(\ref{pdfexp2d}) into Eq.~(\ref{pdfx}) gives 
 \begin{equation}
 p_x(x)=  \frac{\frac{1}{2\sqrt{rD}} e^{-\sqrt{\frac{r}{D}}|x|}+\frac{\sqrt{\pi}x}{4v}G_{1,3}^{3,0}\left(\frac{x^2r}{4D}\left|{\begin{array}{ccc}
   0 \\
   -\frac12 & -\frac12 & \frac12 \\
  \end{array}}\right. \right)}{\frac{1}{r}+\frac{\pi}{2v}\sqrt{\frac{D}{r}}}
 \label{pdfx2d}
 \end{equation}
 with $G_{1,3}^{3,0}$ beeing the Meijer G function. This probability density function $p(x)$ is shown in Fig.~\ref{Gpdf1d}a. The distributions of position projections for non-instantaneous return processes are more strongly peaked at the origin $x=0$ compared to such for instantaneous return. This is due to the fact that a considerable portion of return processes starts around $x=0$ but with large value of $y$, so that the particles stays in the vicinity of $x=0$ during the whole return path.
 
 \begin{figure*}[ht]
\centerline{\includegraphics[width=0.98\columnwidth]{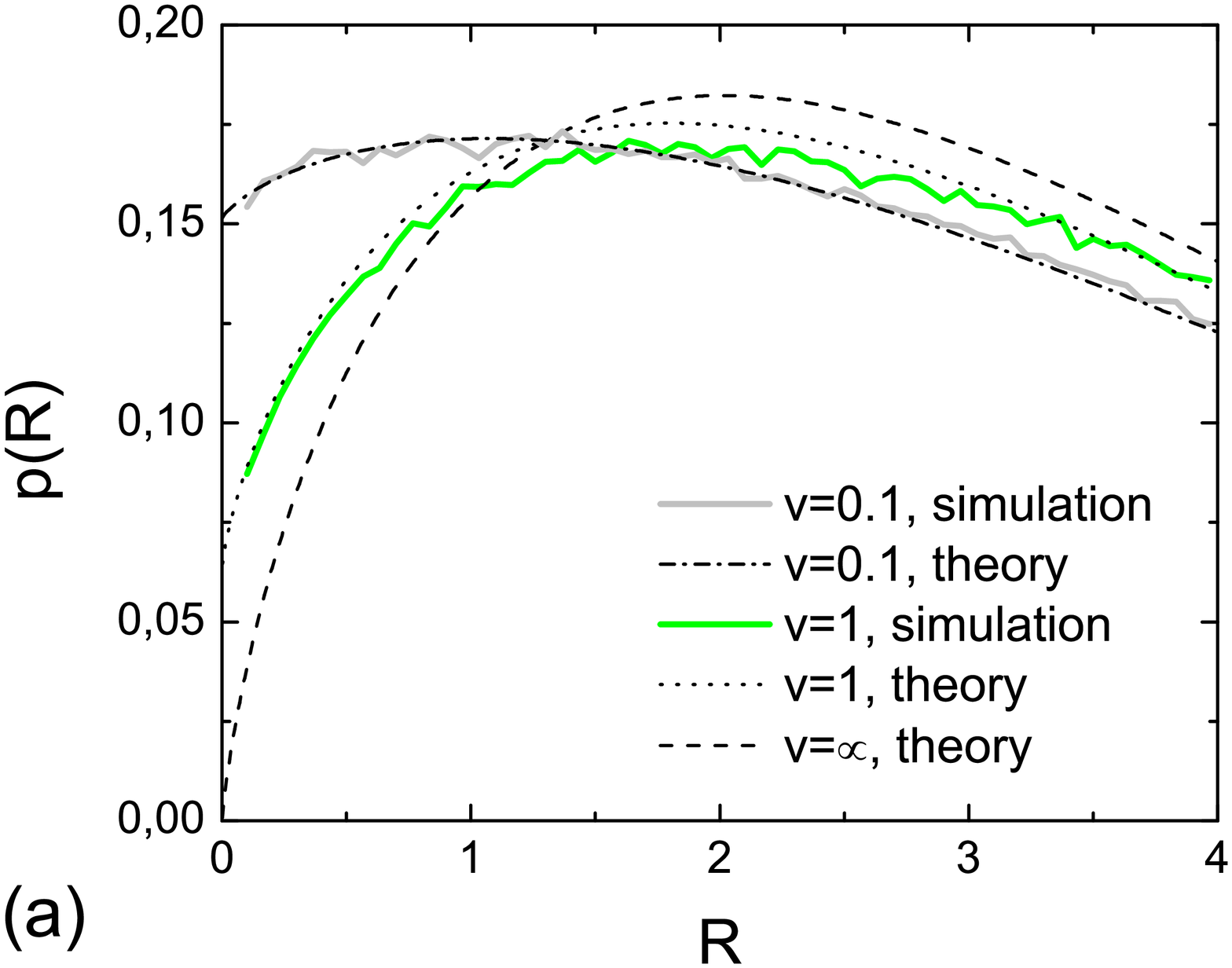}\includegraphics[width=0.98\columnwidth]{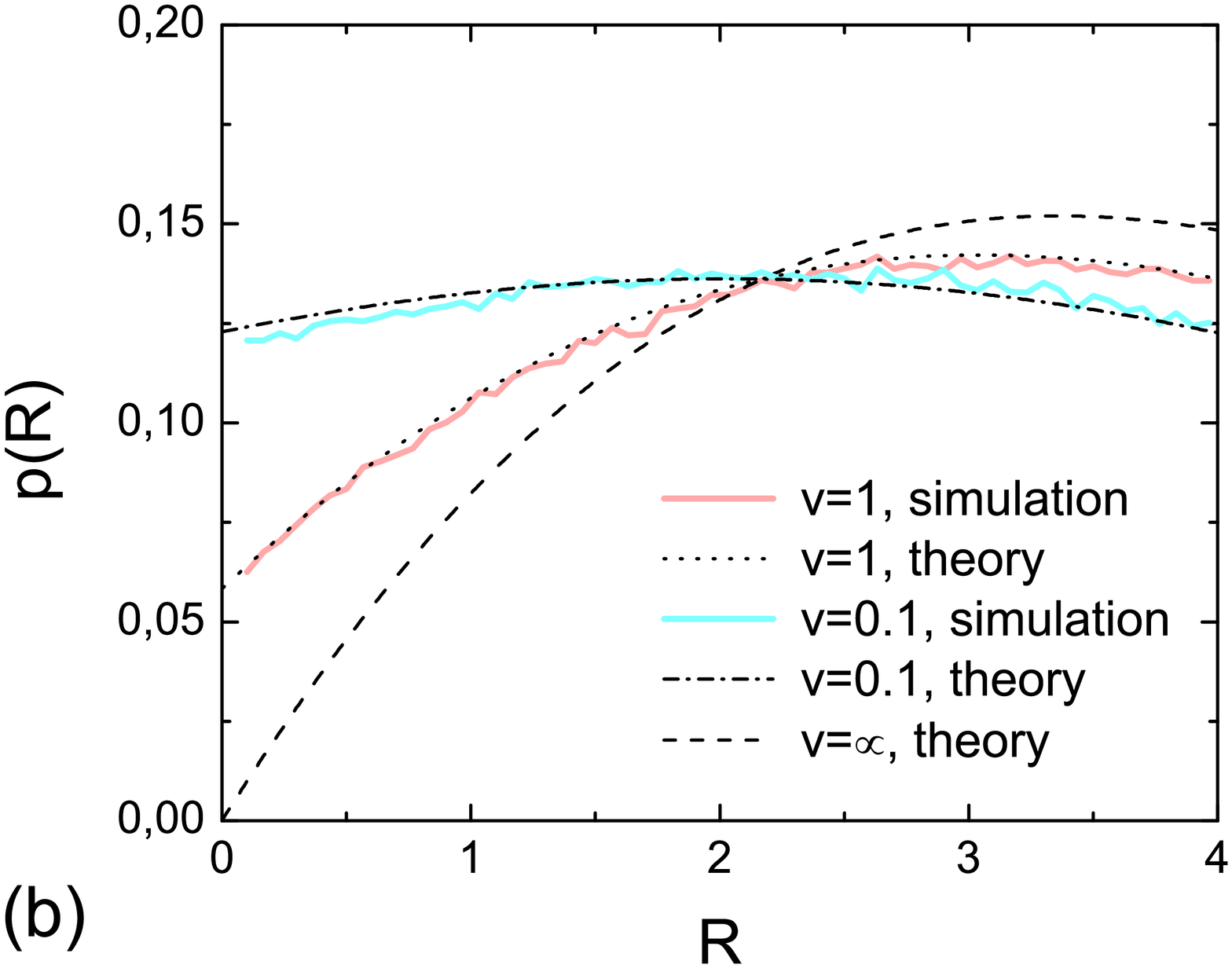}}
\caption{Probability density function for the deterministic resetting in (a) 2d and (b) 3d with $t_r=10$. The colored solid lines show the results of numerical simulations, the thin black dashed and dotted lines show PDF given by Eqs.~(\ref{pdfdet2d}) and~(\ref{pdfdet3d})  for 2d and 3d, respectively. The case $v=\infty$ corresponds to the case of instantaneous resetting.}
\label{Gpdfdet}
\end{figure*} 
 
  \subsubsection{Three dimensions}
The probability density function $p(R)$ in 3d may be obtained by introducing $d=3$ into Eq.~(\ref{rhomainexp})
\begin{equation}\label{pdfexp3d}
p(R)=\frac{\frac{R}{D}e^{-\sqrt{\frac{r}{D}}R}+\frac{1}{v}\left(1+R\sqrt{\frac{r}{D}}\right)e^{-\sqrt{\frac{r}{D}}R}}{\frac{1}{r}+\frac{2}{v}\sqrt{\frac{D}{r}}}\,.
\end{equation}
The PDF given by Eq.~(\ref{pdfexp3d}) is in nice agreement with numerical simulations (see Fig.~\ref{Gexp}b). Again, three-dimensional Brownian motion with Poisson resetting is not invariant with respect to the return velocity.
The MSD in three dimensions is given by
\begin{equation}\label{R2expd3} 
\left\langle R^2\right\rangle = \frac{D}{r}\frac{6v+8\sqrt{Dr}}{v+2\sqrt{Dr}}\,.
\end{equation}
The time evolution of MSD is shown at Fig.~\ref{GR2exp}a. It rapidly tends to the steady state value given by Eq.~(\ref{R2expd3}). This steady state value tends to $\langle R^2 \rangle=6 D/r$ for large velocities (Fig.~\ref{GR2exp}b).

For the distribution of coordinate $x$ we get in three-dimensional systems:
\begin{equation}\label{pdfx3}
p_x(x)= \frac{1}{2}\int_{|x|}^{\infty}dR \frac{p(R)}{R}.
 \end{equation}
 The integration yields
 \begin{equation}
p_x(x)=\frac{\frac{1}{\sqrt{Dr}}e^{-\sqrt{\frac{r}{D}}x}+\frac{1}{2v}e^{-\sqrt{\frac{r}{D}}x}+\frac{1}{2v}\Gamma\left(0;\sqrt{\frac{r}{D}}x\right)}{\frac{1}{r}+\frac{2}{v}\sqrt{\frac{D}{r}}}. \label{pdfx3d}
 \end{equation}
The analytical prediction for $p_x(x)$ is compared with computer simulations at Fig.~\ref{Gpdf1d}b and nice agreement is observed.

\subsection{Deterministic resetting}

\subsubsection{One dimension}
The invariance of the PDF, observed for the normal diffusion with exponential resetting in 1d, does not hold any more for diffusion with deterministic resetting. The PDF may be calculated using Eq.~(\ref{mixture}) \cite{me}:
\begin{equation}\label{pdfdelta}
 p(R)=\frac{\sqrt{\frac{4t_r}{\pi
D}} e^{-\frac{R^2}{4Dt_r}}-\frac{R}{D}
\mbox{erfc}\left(\frac{R}{\sqrt{4Dt_r}}\right)+\frac{1}{v}\mbox{erfc}\left(\frac{R}{\sqrt{4Dt_r}}\right)}{t_r+\frac{1}{v}\sqrt{\frac{4Dt_r}{\pi}}} .
\end{equation}
For the MSD we have
\begin{equation}\label{x2det}
\left\langle R^2\right\rangle = Dt_r\left(1+\frac{\frac{2}{3} \sqrt{Dt_r}}{t_r v\sqrt{\pi}+ 2\sqrt{Dt_r}}\right)    \,.
\end{equation}

\begin{figure*}[ht]
\centerline{\includegraphics[width=0.98\columnwidth]{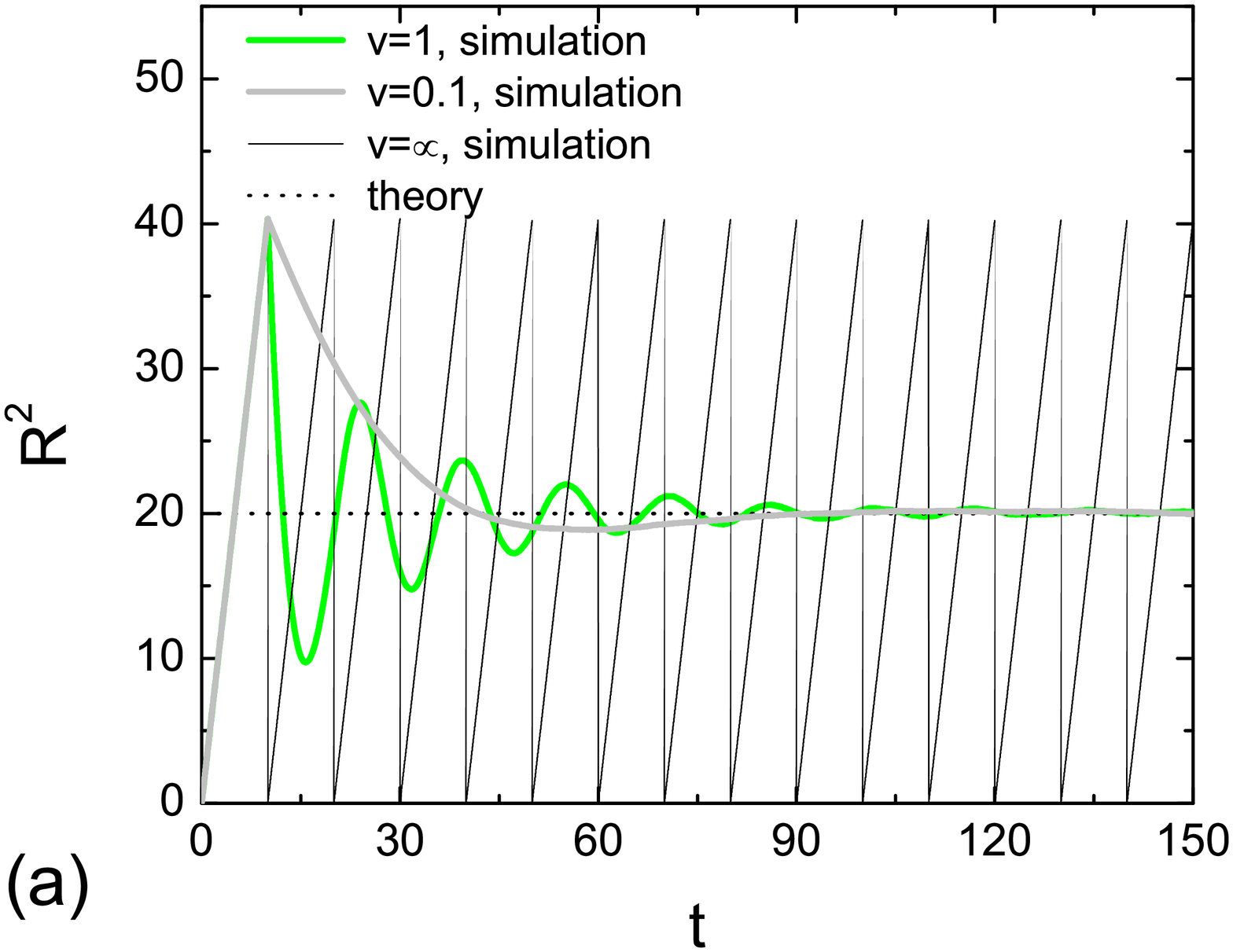}\includegraphics[width=0.98\columnwidth]{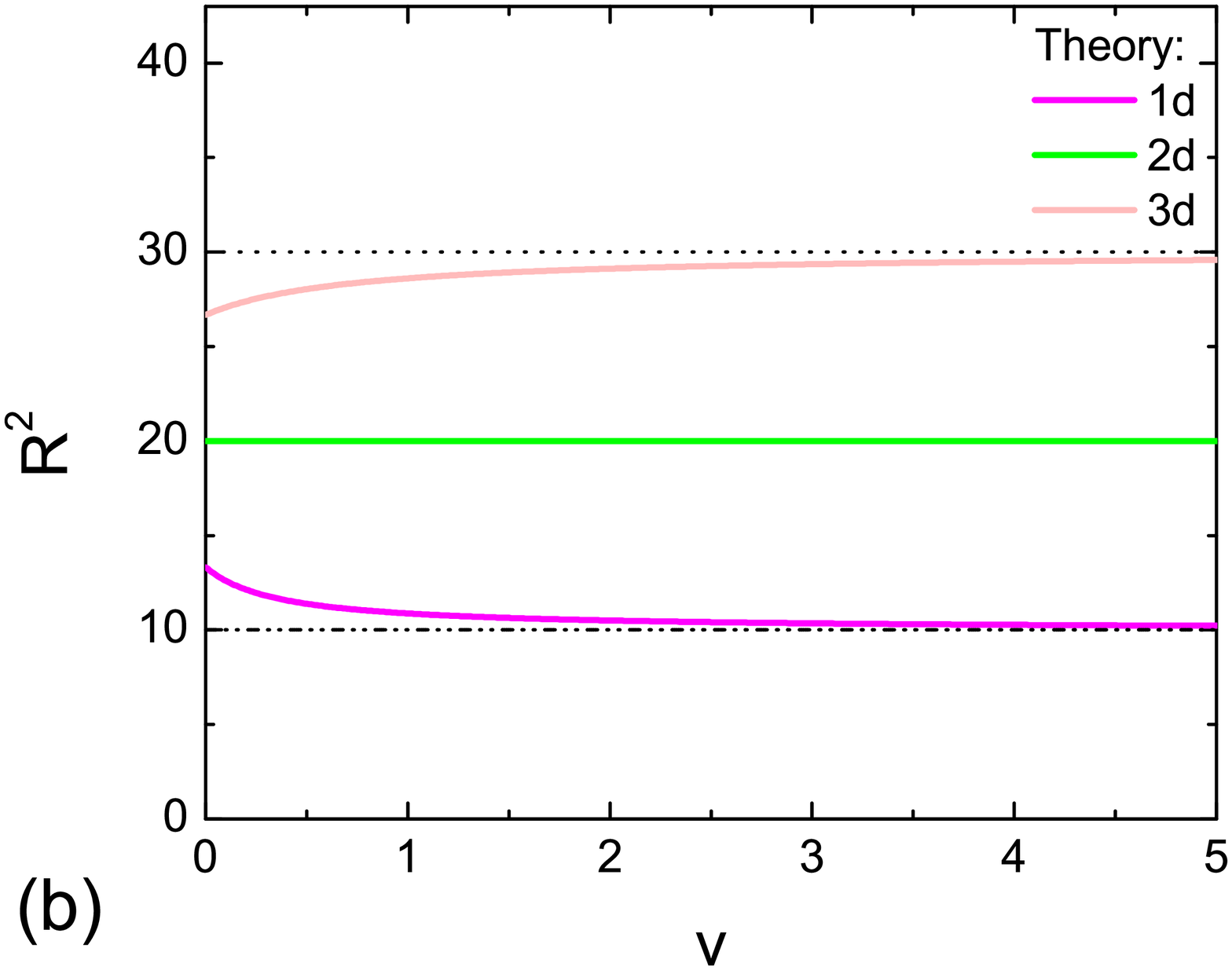}}
\caption{(a) MSD for the deterministic resetting with $t_d=10$ in 2d as a function of time. The MSD oscillates before reaching the steady state given by Eq.~(\ref{R2detd2}). The larger the velocity the longer the oscillation persist. For $v=\infty$ (instantaneous resetting) the oscillations are not damped and persist forever. (b) The dependence of MSD on the return velocity (Eqs.~(\ref{x2det}), (\ref{R2detd2}) and (\ref{R2detd3}) in $1d$, $2d$ and $3d$, correspondingly.) The steady state in 2d is the same for all velocities in contrast to one-and three dimensional systems. }
\label{GR2det}
\end{figure*} 

\subsubsection{Two dimensions}
Using Eq.~(\ref{rhomaindet}) with $d=2$, we get the PDF for two-dimensional Brownian motion with deterministic resetting at constant speed:
\begin{equation}\label{pdfdet2d}
p(R)=\frac{\frac{R}{2D}\Gamma\left(0;\frac{R^2}{4Dt_r}\right)+\frac{1}{v}\exp\left(-\frac{R^2}{4Dt_r}\right)}{t_r+\frac{\sqrt{\pi Dt_r}}{v}}\,.
\end{equation}
Here $\Gamma(0,R)$ is the incomplete Gamma-function. The PDF is shown in Fig.~\ref{Gpdfdet}(a).
The MSD is obtained by substituting $d=2$ into Eq.~(\ref{msdmaindet}):
\begin{equation}\label{R2detd2}
\left\langle R^2\right\rangle =2Dt_r\,.
\end{equation}
While the PDF (Eq.~\ref{pdfdet2d}) explicitly depends on the return velocity, the MSD in the steady state, surprisingly, does not. The effect of the invariance of MSD and PDF with respect to the return velocity will be considered in details in the next Section IV. 

In contrast to the Poisson resetting, when the MSD rapidly reaches the steady state, in the case of the deterministic resetting the MSD oscillates before it reaches the speed-independent stationary value (Fig.~\ref{GR2det}a). The higher the velocity, the longer the oscillations persist. In the case of the instantaneous resetting ($v=\infty$) the oscillations are not damped and persist forever.

The PDF $p_x(x)$ of the position's projection can be derived according to Eq.~(\ref{pdfx}) in full analogy to the case of the exponential resetting:
\begin{eqnarray}\nonumber
&&p_x(x)=\left[\frac{1}{2D}\left(\sqrt{\frac{4Dt_r}{\pi}}e^{-\frac{x^2}{4 D t_r}}-x\,\mathrm{erfc}\left(\frac{x}{\sqrt{4 D t_r}}\right)\right)\right.\\
&&+\left.\frac{1}{2\pi v}\exp\left(-\frac{x^2}{8 D t_r}\right)K_0\left(\frac{x^2}{8 D t_r}\right)\right]/\left(t_r+\frac{\sqrt{\pi Dt_r}}{v}\right)\,.\nonumber\\
\label{pdfxdet2d}
\end{eqnarray}
Here $\rm erfc(R)$ is the complementary error function. $p_x(x)$ is shown at Fig.~\ref{Gpdf1d}(a) and is in a nice agreement with the simulation results.

\subsubsection{Three dimensions}
Substituting $d=3$ into Eq.~(\ref{rhomaindet}), we get the three-dimensional PDF:
\begin{equation}\label{pdfdet3d}
p(R)=\frac{\left(\frac{R}{D}+\frac{1}{v}\right)\mathrm{erfc}\left(\frac{R}{\sqrt{4Dt_r}}\right)+\frac{2R}{v\sqrt{4\pi Dt_r}}\exp\left(-\frac{R^2}{4Dt_r}\right)}{t_r+\frac{4\sqrt{ Dt_r}}{v\sqrt{\pi}}}
\end{equation}
The comparison of this expression to the simulation results is shown in Fig.~\ref{Gpdfdet}(b).
The MSD in 3d attains the form
\begin{equation}\label{R2detd3}
\left\langle R^2\right\rangle =3Dt_r\frac{t_r+\frac{32}{9v}\sqrt{\frac{Dt_r}{\pi}}}{t_r+\frac{4}{v}\sqrt{\frac{Dt_r}{\pi}}}\,.
\end{equation}
The invariance of both PDF and the steady-state MSD with respect to the return velocity is not observed in three-dimensional systems. The dependence of MSD in the steady state on the return velocity $v$ for the deterministic resetting is shown in Fig.~\ref{GR2det}(b). For comparison, also the MSD for one-dimensional system is given. Note that the asymptotic values of the MSD for $v \to \infty$ are in different dimensions are achieved in different ways: The MSD is a
growing function of $v$ in three-dimensional systems, a decreasing function of $v$ in one-dimensional systems, and does not depend on $v$ in two dimensions.

Now let us calculate the PDF of the projections of the locations of particle $p_x(x)$ on $x$-axis. Using Eq.~(\ref{pdfx3}), we have
\begin{eqnarray}\nonumber
&&p_x(x)=\!\!\left[\frac{1}{2D}\!\left(\!\sqrt{\frac{4 D t_r}{\pi}}e^{-\frac{x^2}{4 D t_r}}+x\!\left(\!\mathrm{erf}\left(\!\frac{x}{\sqrt{4 D t_r}}-1\right)\!\right)\!\!\right)\right.\\\nonumber
&&+\,\frac{2}{9v\sqrt{\pi}}\frac{1}{\left(4 D t_r\right)^{3}}x^3\; _2F_2\left(\frac32,\frac32;\frac52,\frac52;-\frac{x^2}{4 D t_r}\right)-\\
&&-\left.\frac{1}{4v}\left(-2+\gamma+\log\left(\frac{x^2}{Dt_r}\right)\right)\right]/\left(t_r+\frac{4\sqrt{ Dt_r}}{v\sqrt{\pi}}\right)
\label{pdfx3det}
 \end{eqnarray}
Here $\rm erf(x)$ is the error function and $_2F_2$ is the generalized hypergeometric function. This function $p_x(x)$ is shown in Fig.~\ref{Gpdf1d}(b) and fits very well the simulation data.

\section{The invariance of the PDF and the MSD with respect to the return velocity}

\subsection{The invariance of the PDF}

As our calculations for the specific examples discussed above show, the PDF of particle's displacements is invariant 
with respect to the return speed in the one-dimensional case for exponential waiting time density, and not invariant
with respect to this speed in two and in three dimensions both for exponential and for the deterministic resetting. 
This motivates a general discussion of circumstances under which the PDF becomes invariant with respect to the return speed. 

Let us return to our Eq.~(\ref{mixture}) and note that according to Eqs.~(\ref{tres}) and (\ref{I1}) $\langle t_{\rm res}\rangle$ and $\rho_1(R)$ are independent on $v$ while both $\rho_2(R,v)$ and in  $\langle t_{\rm ret}(v) \rangle$ are proportional to $v^{-1}$, according to Eqs.~(\ref{I2v}-\ref{tretI}). To stress this velocity dependence 
we have written $v$ as the argument of the corresponding functions explicitly. Now we introduce the function 
$\hat{\rho}_2(R) = v \rho_2(R,v)$ and the parameter $\langle |R_0| \rangle = v \langle t_{\rm ret}\rangle$ 
(having the meaning of the mean position at the end of the displacement phase) which do not depend on $v$
and rewrite our Eq.(\ref{mixture}) in the form
\begin{equation}
 p(R) = \frac{\rho_1(R) + v^{-1}\hat{\rho}_2(R)}{\langle t_{\rm res}\rangle+ v^{-1} \langle |R_0| \rangle}.
\end{equation}
The speed-independence of $p(R)$ then implies that 
\begin{equation}
 \frac{\hat{\rho}_2(R)}{\rho_1(R)}=\frac{\langle |R_0| \rangle}{\langle t_{\rm res}\rangle} = C\,,
\end{equation}
with a constant $C$ not depending neither on $R$ nor on $v$, see Ref.~\cite{me}. 
In our initial notation this means that
\begin{equation}\label{rrc}
\frac{\rho_2(R,v)}{\rho_1(R)}=\frac{\langle t_{\rm ret}(v) \rangle}{\langle t_{\rm res}\rangle} = C.
\end{equation}

The rescaled PDFs in the displacement phase and in the return phase, $\rho_1(R)$  and  $\rho_2(R,v)$, 
Eqs.~(\ref{I2v}) and (\ref{I1}), can be expressed in terms of the function
\begin{equation}
F(R,t)=\int_R^{\infty}p(R^{\prime}|t)dR^{\prime} 
\end{equation}
in the following way \cite{me}:
\begin{eqnarray}
\rho_1(R) = -\int_0^{\infty} \left[\frac{\partial }{\partial R} F(R,t) \right] \Psi(t) dt  \\
\rho_2(R,v) =  \frac{1}{v}\int_0^\infty \left[ \frac{\partial}{\partial t} F(R,t) \right]\Psi(t)dt\,
\end{eqnarray}
with $\Psi(t)$ being the survival probability, Eq.~(\ref{psurv}).
Then the condition given by Eq.~(\ref{rrc}) becomes equivalent to the following integral relation \cite{me}:
\begin{equation}\label{beauty}
 \int_0^\infty \left[C\frac{\partial}{\partial R}F(R,t) +\frac{\partial}{\partial t}F(R,t) \right] \Psi(t)dt =0.
\end{equation}
Now we investigate the question of invariance for normal diffusion in one-, two- and three-dimensional systems.

\subsubsection{One dimension} 
Let us first show that in 1d the exponential waiting time density, Eq.~(\ref{pexp}) is the only solution of integral equation Eq.~(\ref{beauty}) in the case of normal diffusion. The integration of the Gaussian PDF, Eq.~(\ref{pgau}), of a Brownian particle in 1d yields
\begin{equation}
 F(R,t) = \frac{1}{2} \mathrm{erf}\left(\frac{R}{2\sqrt{Dt}} \right).
\end{equation}
Substituting the partial derivatives of $F(R,t)$ into (Eq.~\ref{beauty}), we get
\begin{equation}
\int_0^\infty \sqrt{\frac{1}{t}}\exp\left(-\frac{R^2 }{4Dt} \right) \left(C - \frac{R}{2t} \right) \Psi(t) dt = 0\,.
\end{equation}
Introducing new variables $z=1/t$ and $u = R^2/4D$ we get
\begin{equation}
  \int_0^\infty e^{-uz} \left(\frac{C}{z} - \sqrt{D u} \right) \frac{\Psi\left(1/z\right)}{\sqrt{z}}dz =0\,.
\end{equation}
Introducing a new function 
\begin{equation}
Q(z)= \frac{1}{\sqrt{z}}\Psi\left(\frac {1}{z} \right)
\end{equation}
we rewrite this equation as
\begin{equation}\label{eq:Q}
  C \int_0^\infty \frac{1}{z} e^{-uz} Q(z) dz = \sqrt{Du} \int_0^\infty  e^{-uz} Q(z) dz\,.
\end{equation}
On the r.h.s. we immediately recognize the Laplace transform $\widetilde{Q}(u)$ of $Q(z)$. 
Now we differentiate both parts of this equation with respect to $u$ and rearrange the terms. Thus, 
in the Laplace domain our integral relation turns into the differential equation
\begin{equation}
  \frac{d}{du} \widetilde{Q}(u) = - \left(\frac{1}{2 u} + \frac{C}{\sqrt{Du}} \right) \widetilde{Q}(u) \,.
\end{equation}
This differential equation can be easily integrated
\begin{equation}
 \widetilde{Q}(u) = \frac{C_1}{\sqrt{u}} \exp\left(- \frac{2C}{\sqrt{D}} \sqrt{u} \right),
\end{equation}
with $C_1$ being the constant of integration, and the inverse Laplace transform gives
\begin{equation}
 Q(z) = \frac{\widetilde{C}_1}{\sqrt{\pi z}} \exp\left(-\frac{C^2}{D z}\right)
 \label{eq:SolQ}
\end{equation}
Returning back to $\Psi(t)$ and using the initial condition on the survival probability $\Psi(0)=1$ one gets
\begin{equation}
 \Psi(t) = \exp\left(-\frac{C^2}{D} t \right).
\end{equation}
Introducing the resetting rate $r=C^2/D$ we turn to Eq.~(\ref{survexp}) and therefore obtain exponential resetting as a unique solution for the invariance problem of normal diffusion in one-dimensional systems.

\subsubsection{Two dimensions} 

The integration of the diffusion propagator in 2d, Eq.~(\ref{pgau}), yields
\begin{equation}
 F(R,t) =  1 - \exp\left(-\frac{R^2 }{4Dt} \right).
\end{equation}
The integral relation, Eq.~(\ref{beauty}), becomes
\begin{equation}
  \int_0^\infty \frac{1}{t} \exp\left(-\frac{R^2 }{4Dt} \right) \left( C - \frac{R}{2t} \right) \Psi(t) dt = 0.
\end{equation}
Now we introduce the same variables $z=1/t$ and $u = R^2/4D$ as in 1d, which gives us
\begin{equation}
 \int_0^\infty e^{-uz} \left(\frac{C_2}{z} - \sqrt{D u}\right) \Psi\left(\frac{1}{z} \right)dz =0.
\end{equation}
Denoting 
\begin{equation}
Q(z)= \Psi\left(\frac {1}{z} \right)
\end{equation}
we return to the same Eq.~(\ref{eq:Q}), with the solution given by Eq.~(\ref{eq:SolQ}). In such a way we get
\begin{equation}
 \Psi(t) = \widetilde{C}_2 \sqrt{t} \exp\left(-\frac{C^2}{D} t \right).
\end{equation}
This function, however, vanishes at $t=0$ for any integration constant $\widetilde{C}_2$, therefore the condition $\Psi(0) = 1$ cannot be satisfied, and $\Psi(t)$ cannot be a survival probability. Thus, we have shown, that no physical solution of Eq.~(\ref{beauty}) exists in two dimensions, and the PDF depends on the return velocity $v$ for all possible resetting time distributions.

\subsubsection{Three dimensions} 

In 3d we integrate Eq.~(\ref{pgau}) and obtain
\begin{equation}
 F(R,t) =  \mathrm{erfc}\left(\frac{R}{\sqrt{4Dt}}\right)+\frac{R}{\sqrt{\pi Dt}}e^{-\frac{R^2}{4Dt}}.
\end{equation}
As in previous cases, we obtain Eq.~(\ref{eq:Q}), with the solution given by Eq.~(\ref{eq:SolQ}), but now
with
\begin{equation}
Q(z)= \sqrt{z}\Psi\left(\frac {1}{z} \right)\,.
\end{equation}
The function
\begin{equation}
\Psi(t) = \widetilde{C}_3\, t \exp\left(-\frac{C^2}{D} t \right)
\end{equation}
again can not be considered as a survival probability of the resetting process with any waiting time distribution. So analogously to the two-dimensional systems the PDF depends on the return velocity for any distribution of resetting times.

\subsection{Invariance of the MSD}

Let us consider the question of the invariance of $\langle R^2 \rangle$ with respect to the return velocity. 
The MSD is given by Eq.~(\ref{msdmain}) and depends only on the dimension of space and on four (fractional) moments
of the waiting time distribution. Since such a distribution is not defined by these four parameters uniquely,
there are many waiting time densities, for which an invariance of the MSD with respect to the return speed
could take place. 

\subsubsection{One dimension} 

Using Eq.~(\ref{msdmain}), we get in 1d
\begin{equation}\label{msd1}\langle R^2 \rangle = D\frac{ \langle t^2_{\mathrm{res}}\rangle + \frac{8}{3v}\sqrt{\frac{D}{\pi}} \langle t^{3/2}_{\mathrm{res}}\rangle}{\langle t_{\mathrm{res}}\rangle + \frac{2}{v}\sqrt{\frac{D}{\pi}} \langle t^{1/2}_{\mathrm{res}}\rangle}\,.\end{equation}
The MSD remains invariant if
\begin{equation}  \frac{\langle t^2_{\mathrm{res}} \rangle}{\langle t_{\mathrm{res}}\rangle}= \frac{4}{3} \frac{\langle t^{3/2}_{\mathrm{res}}\rangle}{\langle t^{1/2}_{\mathrm{res}}\rangle}\,,\end{equation}
Using time moments Eqs.~(\ref{tres}-\ref{tres3}), one can see that this condition is fulfilled for the exponential resetting, as expected.

\subsubsection{Two dimensions} 

In two dimensions using Eq.~(\ref{msdmain}), we get 
\begin{equation}\label{msd2}
\langle R^2 \rangle = 2D\frac{ \langle t^2_{\mathrm{res}}\rangle + \frac{\sqrt{\pi D}}{v} \langle t^{3/2}_{\mathrm{res}}\rangle}{\langle t_{\mathrm{res}}\rangle + \frac{\sqrt{\pi D}}{v} \langle t^{1/2}_{\mathrm{res}}\rangle}\,.
\end{equation}
The MSD remains invariant if
\begin{equation}
  \frac{\langle t^2_{\mathrm{res}} \rangle}{\langle t_{\mathrm{res}}\rangle}= \frac{\langle t^{3/2}_{\mathrm{res}}\rangle}{\langle t^{1/2}_{\mathrm{res}}\rangle}\,,
\end{equation}
which holds true for the resetting at fixed time (see Eqs.~(\ref{tresdet}-\ref{tresdeta})).

\subsubsection{Three dimensions} 

Introducing $d=3$ into Eq.~(\ref{msdmain}) we obtain
\begin{equation}\label{msd3}\langle R^2 \rangle = 3D\frac{ \langle t^2_{\mathrm{res}}\rangle + \frac{32}{9v}\sqrt{\frac{D}{\pi}} \langle t^{3/2}_{\mathrm{res}}\rangle}{\langle t_{\mathrm{res}}\rangle + \frac{4}{v}\sqrt{\frac{D}{\pi}} \langle t^{1/2}_{\mathrm{res}}\rangle}\,.\end{equation}
The MSD remains invariant if
\begin{equation}  \frac{\langle t^2_{\mathrm{res}} \rangle}{\langle t_{\mathrm{res}}\rangle}= \frac{8}{9} \frac{\langle t^{3/2}_{\mathrm{res}}\rangle}{\langle t^{1/2}_{\mathrm{res}}\rangle}\,.\end{equation}
To fit a single equation one needs only a one-parametric  family of waiting time distributions with non-negative support. For example, the MSD remains invariant under resetting with the following waiting time distribution:
\begin{equation}
\psi(t)=\frac{1}{2}\left(\delta\left(t-t_0\right)+\delta(t-at_0)\right)\,,
\end{equation}
where the parameter $a$ can be found from the equation
\begin{equation}
\frac{8}{9} a^{5/2} - a^2 - \frac{1}{9}a^{3/2} + \frac{8}{9} - a^{1/2} - \frac{1}{9} = 0
\end{equation}
and is equal either to $a=1.95242$ or to $a=0.442352$.\\

\section{Conclusions}

We have investigated the stationary probability density function (PDF) and the mean-squared displacement (MSD) of Brownian motion under resetting with return at constant velocity in higher dimensions. We have concentrated on particular cases of Poissonian resetting and of resetting at fixed time in one, two- and three dimensional systems. While the stationary PDF 
(and thus the MSD) in one dimension does not depend on the return velocity under exponential resetting, 
no such invariance holds in two or three dimensions. In these cases the probability to find the particle closer to the origin becomes higher and the tails of the PDF become lighter for smaller return velocities. Surprisingly, the steady-state MSD of Brownian motion with deterministic resetting in 2d remains invariant with respect to the return velocity.
 
The problem of the possible invariance of the stationary PDF with respect to return velocity 
was investigated also in a general setting. We have shown that the exponential waiting time distribution 
(characteristic for Poisson resetting) is the only one for which such invariance holds in one dimension. 
We have moreover shown that in two and tree dimensions such an invariance is impossible under any waiting time 
distribution. However, the invariance of the MSD can still be observed.


\begin{thebibliography}{99}

\bibitem{review} M. R. Evans, S. N. Majumdar and G. Schehr, J. Phys. A \textbf{53}, 193001 (2020).
\bibitem{EvansMajumdar} M. R. Evans and S. N. Majumdar, Phys. Rev. Lett. \textbf{106}, 160601 (2011).
\bibitem{high12} M.R. Evans and S.N. Majumdar. J. Phys. A: Math. Theor. \textbf{47}, 285001 (2014). 
\bibitem{bhat} U. Bhat, C. de Bacco, S. Redner. J. Stat Mech. 083401 (2016).
\bibitem{levy1} {\L}. Ku\'smierz, S. N. Majumdar, S. Sabhapandit, and G. Schehr, Phys. Rev. Lett. \textbf{113}, 220602 (2014).
\bibitem{levy2} {\L}. Ku\'smierz, E. Gudowska-Nowak. Phys. Rev. E \textbf{92}, 052127 (2015).
\bibitem{santos} M. A. F. Dos Santos Physics \textbf{1} 40 (2019)
\bibitem{MV2013}  M. Montero and J. Villarroel, Phys. Rev. E \textbf{87}, 012116 (2013).
\bibitem{MC2016} V. M$\rm\acute{e}$ndez and D. Campos, Phys. Rev. E \textbf{93}, 022106 (2016).
\bibitem{Sh2017}  V.P. Shkilev, Phys. Rev. E \textbf{96}, 012126 (2017).
\bibitem{ctrw} M. Montero, A. Mas-Puigdell\'osas, J. Villarroel. Eur. Phys. J. B \textbf{90}, 176 (2017).
\bibitem{ctrwour} A.S. Bodrova, I.M. Sokolov. Phys. Rev. E \textbf{101}, 062117 (2020).
\bibitem{rap} P. Singh. https://arxiv.org/abs/2007.05576
\bibitem{Anna01} A.S. Bodrova, A.V. Chechkin, I.M. Sokolov. Phys. Rev. E \textbf{100}, 012119 (2019).
\bibitem{Anna02} A.S. Bodrova, A.V. Chechkin, I.M. Sokolov. Phys. Rev. E \textbf{100}, 012120 (2019). 
\bibitem{shlomi2017} A. Pal and S. Reuveni. Phys. Rev. Lett. \textbf{118}, 030603 (2017).
\bibitem{palrt} A. Pal, A. Kundu and M. R. Evans, J. Phys. A: Math. Theor. \textbf{49}, 225001 (2016).
\bibitem{NagarGupta} A. Nagar and S. Gupta, Phys. Rev. E \textbf{93}, 060102(R) (2016).
\bibitem{res2016} S. Eule and J. J. Metzger, New J. Phys. \textbf{18}, 033006 (2016).
\bibitem{shlomi2016} S. Reuveni. Phys. Rev. Lett. \textbf{116}, 170601 (2016).
\bibitem{computerscience}  A. Montanari and R. Zecchina, Phys. Rev. Lett. \textbf{88}, 178701 (2002).
\bibitem{shlomi} A. Pal, {\L}. Ku\'smierz, S. Reuveni, https://arxiv.org/abs/1906.06987.
\bibitem{shlomi1} A. Pal, {\L}. Ku\'smierz, S. Reuveni, New J. Phys. \textbf{21}, 113024 (2019).
\bibitem{shlomi2} A. Pal, {\L}. Ku\'smierz, S. Reuveni, Phys. Rev. E \textbf{100}, 040101(R) (2019).
\bibitem{campos} A. Mas$\rm\acute{o}$-Puigdellosas, D. Campos, and V. M$\rm\acute{e}$ndez, Phys. Rev. E \textbf{100}, 042104 (2019).
\bibitem{me} A.S. Bodrova and I.M. Sokolov. Phys. Rev. E. \textbf{101}, 052130 (2020).
\bibitem{exp} O. Tal-Friedman, A. Pal, A. Sekhon, S. Reuveni, Y. Roichman. https://arxiv.org/abs/2003.03096
\bibitem{exp2} B. Besga, A. Bovon, A. Petrosyan, S. N. Majumdar, S. Ciliberto.
https://arxiv.org/abs/2004.11311
\bibitem{sokbook} J. Klafter and I.M. Sokolov, First Steps in Random Walks: From Tools to Applications. Oxford University Press, New York, USA (2011).
\bibitem{sokprl} A.V. Chechkin, I.M. Sokolov, Phys. Rev. Lett. \textbf{121}, 050601 (2018).

\end{thebibliography}
 \end{document}